\documentclass[aps,twocolumn,superscriptaddress]{revtex4-1}

\usepackage[colorlinks]{hyperref}
\usepackage{graphicx}
\usepackage{amsmath}
\usepackage{amsfonts}
\usepackage{mathrsfs}
\usepackage{color}

\newcommand{\bs}[1]{\boldsymbol #1}

\newcommand{\bge}{\begin{equation}}
\newcommand{\ene}{\end{equation}}
\newcommand{\bgea}{\begin{eqnarray}}
\newcommand{\enea}{\end{eqnarray}}

\begin{document}
\date{\today}  

\title{Ultrafast cooling and heating scenarios for the laser induced phase transition in CuO}
\author{Johan Hellsvik}
\email[]{hellsvik@kth.se}
\affiliation{Department of Materials and Nano Physics, KTH Royal Institute of Technology, Electrum 229, SE-164~40 Kista, Sweden}
\affiliation{Dipartimento di Fisica, Universit\`a di Roma ``La Sapienza'', P.  Aldo Moro 2, 00185 Roma, Italy}
\affiliation{Istituto dei Sistemi Complessi, Consiglio Nazionale delle Ricerche, Italy}
\author{Johan H. Mentink}
\affiliation{Radboud University Nijmegen, Institute of Molecules and Materials, Heyendaalseweg 135, 6525 AJ Nijmegen, The Netherlands}
\affiliation{Max Planck Research Department for Structural Dynamics, University of Hamburg-CFEL, 22761 Hamburg, Germany}
\author{Jos\'{e} Lorenzana}
\affiliation{Dipartimento di Fisica, Universit\`a di Roma ``La Sapienza'', P.  Aldo Moro 2, 00185 Roma, Italy}
\affiliation{Istituto dei Sistemi Complessi, Consiglio Nazionale delle Ricerche, Italy}

\begin{abstract}
{The multiferroic compound CuO exhibits low temperature magnetic properties similar to antiferromagnetic iron oxides, while the electronic properties have much more in common with the high $T_c$ cuprate superconductors. This suggests novel possibilities for the ultrafast optical excitation of magnetism. On the basis on atomistic spin dynamics simulations, we study the effect of phonon-assisted multimagnon absorption and photodoping on the spin dynamics in the vicinity of the first-order phase transition from collinear to spin-spiral magnetic order. Similar as in recent experiments, we find that for both excitations the phase transition can proceed on the picosecond timescale. Interestingly, however, these excitation mechanisms display very distinct dynamics. Following photodoping, the spin system first cools down on sub-ps timescales, which we explain as an ultrafast magnetocaloric effect. Opposed to this, following phonon-assisted multimagnon excitation the spin systems rapidly heats up and subsequently evolves to the noncollinear phase even under the influence of isotropic exchange interactions alone.}
\end{abstract}

%\keywords{}

\maketitle

\section{Introduction}
The exploration of the speed limits of phase transitions between macroscopically ordered phases of condensed matter systems has evolved into a central branch of physics, including the study of magnetic phase transitions \cite{Kirilyuk2010,Radu2011,Li2013a}, metal-insulator transitions \cite{Rini2007,Aetukuri2013,DeJong2013,Borroni2015v1} and light-induced superconductivity \cite{Fausti2011,Mankowsky2014}. Moreover, recently, this investigation has even been extended to multiferroic systems, which already in the ground state can exhibit both electric and magnetic order and the coupling between them can give rise to highly non-trivial dynamics, such as large amplitude spin-cycloid rotation \cite{Kubacka2014}, colossal dynamical magnetoelectric effect \cite{Takahashi2012}, and optically driven ultrafast transition from collinear antiferromagnetic (AFM) to spiral states in the multiferroic CuO \cite{Johnson2012}. 

Gaining microscopic understanding of laser-induced ultrafast phase transitions is in general a highly challenging computational problem since it requires the modeling of the effect of laser absorption, which implies the simulation of coupled time-evolution of the charge, spin and lattice degrees of freedom. Nevertheless, important insights can be obtained by breaking this problem into smaller parts. For example, laser-induced demagnetization in ferromagnetic (FM) metals was successfully modelled by assuming that the laser pulse heats the electrons far above the Curie temperature on femtosecond (fs) timescales \cite{beaurepaire1996,koopmans2005,Atxitia2007,kazantseva2008,koopmans2010,evans2015}. Moreover, when different FM metals are coupled antiferromagnetically, even laser-heating alone was found to be sufficient to cause spin switching \cite{Radu2011,Mentink2012,Ostler2012}. On the other hand, the vast majority of intrinsic antiferromagnetic materials are insulating, and for excitation below the charge transfer gap such ultrafast electronic heating is not possible. Nevertheless, ultrafast laser-induced spin-reorentation phase transitions have been observed in iron oxides, for example, and were explained by the coexistence of two metastable spin-orientation phases enabling a spin-inertia driven spin switching mechanism \cite{Kimel2009,Afanasiev2016} and by laser-induced heating of the phonons \cite{Kimel2004,DeJong2012}.

Here we focus on a different type of spin-reorentation transition: the transition from the collinear AFM phase to the spin-spiral phase in CuO. Opposed to iron oxides, the electronic structure of CuO has much in common with the cuprate high-temperature superconductors, yielding different possible excitation mechanisms. First, in these materials it is well-known that the optical absorption below the charge transfer gap is dominated by phonon-assisted multimagnon (MM) excitation \cite{Lorenzana1995,Lorenzana1995a}. Second, for these Mott-insulating materials it was recently predicted that photo doping (PD) can cause an ultrafast modification of the exchange interaction, with an effect comparable to chemical doping \cite{Mentink2014}. The latter is particularly interesting since recent investigations show that chemical doping dramatically reduces the temperature at which the antiferromagnetic to spin-spiral phase in CuO occurs \cite{Hellsvik2014}. Hence, both MM excitation and PD suggest distinct and rather direct ways to induce this phase transition, which so far have not been explored. 

In this article we investigate systematically phonon-assisted MM excitation and PD as mechanisms for the laser-induced phase transition in CuO on the basis of atomistic spin dynamics (ASD) simulations. We demonstrate that both excitation mechanisms can induce the phase transition on picosecond (ps) timescales. Intriguingly, we find that different excitation mechanisms nevertheless induce qualitatively very different dynamics in which the spin system either cools down or heats up on sub ps timescales. In addition, we find that the fastest phase transition occurs for phonon-assisted MM excitation and demonstrate that only in this case the phase transition proceeds even in the absence of anisotropy and dissipation. This suggest that exchange interactions between the spins alone are sufficient to drive the system across the transition, which explains why it can be faster than the dynamics caused by PD.

The paper is organized as follows: The properties of CuO are discussed in Sec.~\ref{sec:cuo}. The model for CuO and the semiclassical equation of motion method are introduced in Sec.~\ref{sec:model}. The different  excitation mechanisms are treated in Sec.~\ref{sec:excmech} and we discuss how they are modelled within the ASD simulations. The results from such simulations are presented in Sec.~\ref{sec:sims} and compared with the results reported from ultrafast time resolved pump-probe measurements~\cite{Johnson2012}. A discussion and outlook is given in Sec.~\ref{sec:conc}.

\begin{figure}
\includegraphics[width=0.50\textwidth]{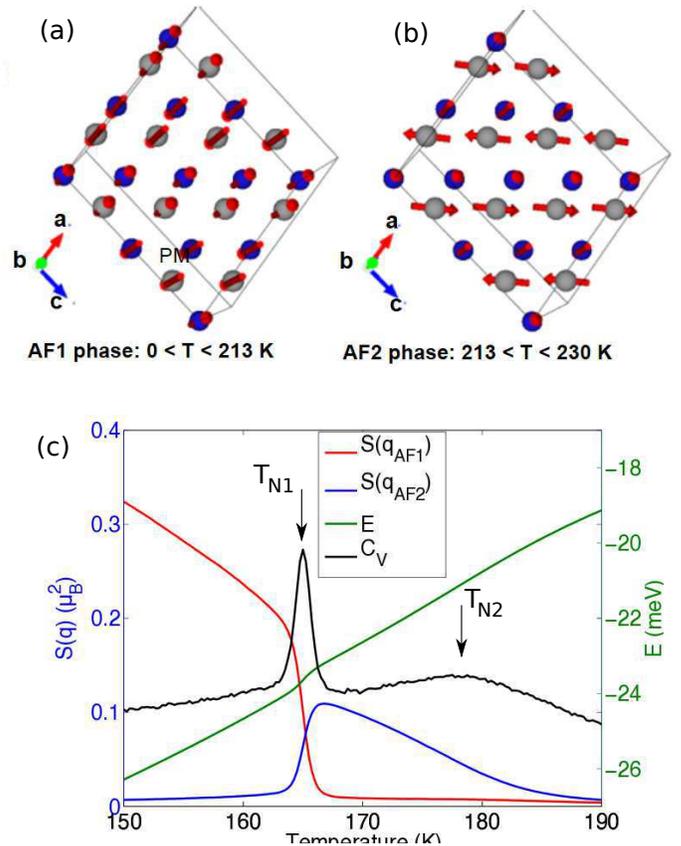}
\caption{\label{fig:PhaseDiagramCuO}The spin configurations of the (a) AF1 and (b) AF2 phases in CuO. Blue and grey spheres are Cu atoms on different $(010)$ planes \cite{Giovannetti2011,Hellsvik2014}. The red arrows show the Cu spin polarization. Entering the AF2 phase, the $\mathbf{q}_{\rm AF1}=(0.5, 0, -0.5)$ ordering vector is modulated to the incommensurate ordering vector $\mathbf{q}_{\rm AF2}=(0.506, 0, -0.483)$. (c) The static structure factor $S(\mathbf{q})$ for $\mathbf{q}_{\rm AF1}$ and $\mathbf{q}_{\rm AF2}$, and the total energy $E$ and heat capacity $C_V$ per spin, obtained in classical Monte Carlo simulations for a $36\mathbf{a}\times 4\mathbf{b}\times 36\mathbf{c}$ supercell using the parameters of Ref.~\cite{Hellsvik2014}. The structure factor is normalised to unity for the $T=0$~K AF1 spin configuration.}
\end{figure}

\section{Magnetic properties of CuO}\label{sec:cuo}

CuO is a Mott insulator and spin $S=1/2$ compound with a strongly
dominant exchange interaction along the crystallographic $[10\bar{1}]$
axis. At high temperatures it is effectively a one-dimensional quantum
spin chain with fermionic spinon excitations \cite{Jung2009}. Below a
weakly first order transition at $T_{N2}=230$ K, a three dimensional
magnetic ordering (AF2) sets in, which over a 17 K temperature span
down to $T_{N1}=213$ K, is an incommensurate spin ordering with
nearest neighbour spins at nearly 90 degrees to each other
[Fig.~\ref{fig:PhaseDiagramCuO}(b)]. This phase has been identified as
multiferroic by Kimura and collaborators \cite{Kimura2008}.

Below the first order phase
transition at $T_{N1}$, a commensurate collinear ordering (AF1) is
found [Fig.~\ref{fig:PhaseDiagramCuO}(a)]. The ordering of the AF1 and
AF2 phases and the overall temperature dependence of the phase diagram
[Fig.~\ref{fig:PhaseDiagramCuO}(c)] can be understood in terms of a competition between the frustration among the Heisenberg couplings, which promotes the AF2 configuration, with the biquadratic exchange couplings, which suppresses the non-collinear ordering for temperatures up to $T_{N1}$ \cite{Yablonskii1990,Hellsvik2014}.

The unusually high temperature of the AF2 multiferroic phase has been
a strong driving force for experimental and theoretical activity
\cite{Giovannetti2011,Jin2012,Toledano2012b,Rocquefelte2013,Villarreal2012,Staub2014,Jones2014,Jones2014a,Cao2015}
on cupric oxide, CuO. In this phase the canted magnetism drives
improper ferroelectricity\cite{Giovannetti2011,Jin2012,Toledano2012b,Rocquefelte2013,Villarreal2012,Hellsvik2014}. Whereas the understanding of the
equilibrium phase diagram has progressed, it is unclear how the phases
evolve into each other upon laser excitation. 

Using a free electron laser as source for the probe beam, time-resolved diffraction experiments revealed an ultrafast optically driven phase transition in CuO from the AF1 phase to the AF2 phase \cite{Johnson2012}. The diffraction signal gave immediate insight into the evolution in time of the magnetic ordering upon excitation. Measurements were performed at $T=207$ K, slightly below $T_{N1}$. The process was fluency-dependent up to a saturation fluency for which a minimum transition time of 400 fs was reported. In the next section we discuss the model and methods we use to simulate the dynamics between these two phases.

\section{Model and methods}\label{sec:model}
The phase competition between the AF1 and AF2 phases in CuO can be captured by a classical spin Hamiltonian which includes the competition between bilinear Heisenberg exchange and biquadratic exchange \cite{Yablonskii1990,Villarreal2012,Hellsvik2014}.

In recent theoretical studies, density functional theory calculations have been pursued to parametrise these exchange interactions from the total energies for different spin configurations \cite{Filippetti2005,Jin2012,Rocquefelte2013,Hellsvik2014}. For the present investigations the Hamiltonian  
\begin{eqnarray}\label{eq:hmag}
&& \mathscr{H}_{\mathrm{M}} = \mathscr{H}_{\mathrm{exch}} +
\mathscr{H}_{\mathrm{bq}} + \mathscr{H}_{\mathrm{ani}} \\
&=&\frac{1}{2}\sum_{i\neq j}J_{ij}\mathbf{S}_i\cdot\mathbf{S}_j +
\frac{1}{2}\sum_{i\neq j}K_{ij}(\mathbf{S}_i\cdot\mathbf{S}_j)^2 +
\frac{1}{2}\sum_{i\neq j} \mathbf{S}_i \mathbf{J}_{ij}^{\mathrm{ani}}
\mathbf{S}_j \nonumber
\end{eqnarray}
is used, where the first term is Heisenberg exchange (exch), the second term is biquadratic exchange (bq) and the third term is anisotropic exchange (ani) and $\mathbf{S}_i$ is a classical spin of magnitude $|\mathbf{S}_i|=S_i=1/2$.

With the set of parameters reported in Ref.~\cite{Hellsvik2014}
classical Monte Carlo simulations of Eq.~\ref{eq:hmag} produce a phase
diagram with a first order phase transition at $T_{\rm N1}\approx
165$~K, and a weakly first order phase transition to the paramagnetic
phase at $T_{\rm N2}\approx 179$~K. In Figure
\ref{fig:PhaseDiagramCuO}(c) is shown the structure factor
$S(\mathbf{q})$ for the ordering vectors $\mathbf{q}_{\rm AF1}=(0.5,
0, -0.5)$ and $\mathbf{q}_{\rm AF2}=(0.528, 0, -0.472)$ in a
temperature interval around the phase transition at $T_{\rm N1}$. To
go from the commensurate AF1 phase to the incommensurate AF2 phase
latent heat needs to be supplied represented by the sharp peak in the
specific heat. Except for the underestimation of the N\'eel temperatures,
the phase diagram is in excellent agreement with experiment for the
nature of the phases, wave-vector and nature of the order
parameters, etc \cite{Hellsvik2014}. 

Given the good agreement for the equilibrium phase diagram, we will use ASD simulations based on the Hamiltonian Eq.~\eqref{eq:hmag} as minimal approach for studying the nonequilibrium spin dynamics across the AF1-AF2 phase transition. Within ASD, the dynamics of a spin system is modelled semiclassically by the stochastic Landau-Lifshitz equation (SLL)
\begin{equation}
\hbar\frac{d\mathbf{S}_i}{dt}=- \mathbf{S}_i \times [\mathbf{B}_{i}+\mathbf{B}_{i}^{\mathrm{fl}}(t)]-\frac{\alpha}{S_i} \mathbf{S}_i \times \{\mathbf{S}_i \times [\mathbf{B}_{i}+\mathbf{B}_{i}^{\mathrm{fl}}(t)]\},
\label{eq:sllg}
\end{equation}
which describe the motion of the classical spins $\mathbf{S}_i$ in an
effective magnetic field $\mathbf{B}_{i}$, calculated from
$\mathbf{B}_i=-\frac{\partial \mathscr{H}}{\partial
  \mathbf{S}_i}$. $\mathbf{B}_{i}^{\mathrm{fl}}(t)$ is a stochastic
magnetic field with a Gaussian distribution. By the
fluctuation-dissipation theorem, the magnitude of
$\mathbf{B}_{i}^{\mathrm{fl}}(t)$ is related to the dimensionless damping parameter 
$\alpha$. 
 
We note that the dynamics we investigate are
at temperatures high enough to allow for the assumption of a classical
renormalised regime \cite{Chakravarty1989} which renders
Eq.~(\ref{eq:sllg}) applicable even for a spin $1/2$ system. 

To characterise the dynamics across the phase transition we focus on two observables. First, we compute the equal time correlation function $S(\mathbf{q},t)=\frac{1}{2\pi} \sum_i e^{\text{i}\mathbf{q}\cdot\mathbf{r}_i}C(\mathbf{r}_i,t)$ where $C(\mathbf{r}_i-\mathbf{r}_j,t)=[\mathbf{S}_i(t)\cdot\mathbf{S}_j(t)]$ is the spin-spin correlation function for spins displaced by the vector $\mathbf{r}_i-\mathbf{r}_j$. $[\ldots]$ indicate averaging over the cell and over different realisations of the initial spin excitations and the heat bath. $S(\mathbf{q},t)$ shows two peaks at $\mathbf{q}=\mathbf{q}_{\rm AF1}$ and $\mathbf{q}=\mathbf{q}_{\rm AF2}$ which we use to monitor the evolution from one phase to the other. 

Second, we monitor the evolution of the temperature of the spin system. To this end, we sample the energy $E_i$ defined as the difference
 \begin{eqnarray}
E_i &=& E-E_i^g, 
\end{eqnarray}
between the instantaneous total energy $E$ and the total energy $E_i^g$ when spin $\mathbf{S}_{i}$ is relaxed keeping all other spins fixed. The energy $E_i^g$ is minimised, for fixed orientation of the other spins $\mathbf{S}_{j\neq i}$, with a gradient descent algorithm. The temperature $T$ is then obtained by fitting the histogram of $\{E_i\}$ to the Boltzmann distribution
\bgea\label{eq:boltz}
P[E_i]=P_0e^{-\beta E_i},
\enea
where $k_B$ is the Boltzmann constant and $\beta=1/(k_BT)$ is the inverse temperature. The method of fitting the histograms of $\{E_i\}$ to a Boltzmann distribution assumes that the spins are close to the local equilibrium state and is therefore only applicable for moderate temperatures. We have investigated the range of applicability in more detail in Appendix \ref{sec:tempmeas}, including a quantitative comparison with an alternative approach for computing the spin temperature \cite{Ma2010}, and found our approach to be sufficiently accurate to capture the temperature regime studied below. 

\section{Excitation mechanisms}\label{sec:excmech}
In general, the light-matter interaction in the optical regime
  relies on the electric dipole approximation. This makes the direct
  excitation of the magnetic degrees of freedom in materials with
  inversion symmetry weak or even forbidden. In addition, different
  from magnetic metals, in which the rapid heating of the electrons
  turns out to be very efficient for driving (sub) picosecond phase
  transitions \cite{Mentink2012,Ostler2012}, CuO is insulating and
  photon absorption does not seem to be relevant on first
  sight. Below, however, we discuss three distinct mechanisms
  which indirectly couple light to the magnetic degrees of freedom. All
  mechanisms rely, on one way or the other, on modulation of exchange
  interactions $J$, either by joint excitation of a phonon (phonon
  assisted infrared absorption), by changing
  the electronic distribution (photodoping) or by modulating $J$ with
  an electric field (two-magnon Raman scattering). 
We then estimate their efficiency based on literature values and explain how we effectively include these excitation mechanisms within atomistic spin dynamics simulations. 

\subsection{Infrared absorption}

In materials with inversion symmetry, such as CuO in the AF1 phase, the excitation of magnons by absorption of infrared photons is in principle electric dipole forbidden. However, the inversion symmetry can be broken by the excitation of a phonon which enables simultaneous excitation of an even number of magnons \cite{Lorenzana1995,Lorenzana1995a}. This mechanism dominates for infrared absorption in CuO layers and similar magnetically ordered insulators and is described by the Hamiltonian:
\bge \label{e:spinph}
H =
\sum_{i,{\delta}}J(\bs{E},\mathbf{u}_{i+{\delta}/2})\bs{S}_i\bs{S}_{i+{\delta}}+
\bs{P}_{ph}\cdot\bs{E}
 + H_{ph} 
\ene
Here $\bs{E}$ describes the electric field of light, $\bs{u}_i$ the
phonon displacements and $J(\bs{E},\mathbf{u}_{i+{\delta}/2})$ the
exchange interaction modified by the electric field and the phonon
displacements. Further $\delta$ indicates nearest neighbor sites with
respect to site $i$, while $\bs{P}_{ph}= \sum_{i,{\delta}} q_0 \mathbf{u}_{i+{\delta}/2} $ and $H_{ph}$ are the phonon dipole moment and phonon Hamiltonian, respectively. 

Expanding the Hamiltonian to linear order in the electric field and
the ionic displacement one obtains the dipole moment for Cartesian
component $\mu$, 
\begin{eqnarray}
  \label{eq:p}
  P_{\mu} &=& P_{ph,\mu}+ \frac{\partial J}{\partial  E_\mu}\bs{S}_i\bs{S}_{i+{\delta}}  \\
&+&\sum_{i,{\delta},\nu} \frac{\partial J}{\partial  E_\mu \partial
  u_{i+{\delta}/2,\nu}} 
u_{i+{\delta}/2,\nu}\bs{S}_i\bs{S}_{i+{\delta}}\nonumber
\end{eqnarray}
 Neglecting the small ferroelectric effects in the multiferroic phase
the second term is zero in the presence of inversion symmetry. 
The third term is the one responsible for the phonon-assisted absorption of magnons. 

We observe that the phonon breaks the inversion symmetry as required
for non-vanishing electric-dipole transitions. We note that opposed to
the AF1 phase, the AF2 phase exhibits a ferroelectric dipole moment
which in principle facilitates a direct coupling of the electric field
component of the laser pulse to the magnetic degrees of
freedom through the second term of Eq.~\eqref{eq:p}. However, such coupling is only effective once the system is already in the AF2 phase, whereas we want to address how fast the
reorientation of the spin system from the collinear AF1 phase to the
incommensurate spin spiral AF2 phase can occur. In any case, we expect
spontaneous inversion symmetry  effects to be much smaller since the involved dipoles are
minute. Indeed, they are beyond the sensibility
of X-ray experiments, for example, and are only detected through the
macroscopic polarization.  

Clearly, it would be very interesting to study the complete
  dynamics of the spin and lattice degrees of freedom that follow from
  Eq.~\eqref{e:spinph}. This, however, is very challenging since it
  requires at least to go beyond the semi-classical description for
  the spin dynamics. In addition, augmenting classical phonon dynamics
  and their coupling to ASD would only slightly change the spin
  dynamics in the regard that displacement (spin) correlations
  renormalize the magnon (phonon) frequencies, but this spin-phonon
  coupling is not strong enough to change the order of time-scales of
  the transversal spin reorientation. Therefore, here we focus
  exclusively on the classical spin dynamics following phonon-assisted
  MM excitation and omit solving directly the time evolution during
  the excitation of magnons and phonons.

To effectively take this excitation into account we assume each
absorbed photon creates a phonon plus multiple two-magnon excitations. We
neglect the first and simulate the second by an exchange process. Namely, we interchange a fraction of nearest neighbor spins along the $[10\bar{1}]$-direction, which has the dominant exchange coupling in CuO \cite{Jung2009,Hellsvik2014}. To estimate the fraction of spin pairs that are interchanged, we use that in middle-IR absorption measurements with photon energies below the charge transfer gap, the absorption coefficient of CuO was determined to be of the order $\alpha\sim$ 200 cm$^{-1}$ (Ref.~\cite{Moskvin1994}), with corresponding skin penetration depth $d\sim 0.005$ cm. Defining the average number of absorbed photons per unit volume $N_{\rm phot}$, the number $\rho$ of absorbed photons per Cu atom can be computed through the expression
\bgea
\rho&=&N_{\rm phot}V_{\rm Cu}=\frac{E_{\rm pump}}{E_{\rm phot}}\frac{1}{d}V_{\rm Cu}
\enea
with $\rho$ taking values $4.6\cdot 10^{-4}$ $(1.8\cdot 10^{-3} )$ for laser fluency 7 mJ/cm$^2$ (28 mJ/cm$^2$). Each absorbed photon cause multiple phonon assisted magnons pairs \cite{Perkins1998,Lorenzana1995,Lorenzana1995a}. Therefore, in the simulations that are presented in Sec.~\ref{sec:sims}, a fraction in the range $[x=0.005, \ldots 0.040]$ of randomly chosen pairs of spins are interchanged. This yields a small but finite probability also for four magnon scattering which take place when two pairs of exchanged spins are in the immediate vicinity of each other. 

\subsection{Photo doping}
Photon absorption at frequencies comparable to the gap can cause PD, in which the electronic distribution is changed, which subsequently can modify the exchange couplings and spin ordering \cite{Li2013a,Mentink2014}. For Mott-insulators like CuO, PD comprises the creation of holon-doublon pairs (h-d pairs) and their relaxation within the Hubbard bands causes a rapid reduction of the ordered spin moment $\langle \mathbf{S}_i\rangle$ and the exchange parameter within a few tens of the hopping time \cite{Mentink2014}. Since electron hopping occurs on the fs time scale, PD has the potential of being fast enough for launching spin dynamics that lead to the phase transition on (sub) picosecond time scales. In a classical picture, a simplified way to account for the presence of h-d pairs is by introducing vacancies. Alternatively, one can introduce reduced local moments [similarly to a reduction of $S(\mathbf{q}_{\rm AF1},t)$], in conjunction with reduced exchange parameters. For small concentrations $x$ we have checked these approaches produce qualitatively very similar results. In the simulation results presented below, we have therefore restricted ourselves to the introduction of vacancies. Note that, we could also let the next nearest neighbor (NNN)
couplings of the vacancy be changed, in line with the second order Henley effect in Ref.~\cite{Hellsvik2014}, but for simplicity we limited ourselves to the introduction of vacancies alone. In addition, we assume that the vacancies persist over time scales
longer than the ASD simulation time ($t>5$ ps), in order to isolate
purely the effect of PD alone. For the concentration of vacancies,
values in the range $[x=0.001, \ldots 0.050]$ have been used (c.f. the
case of phonon assisted MM excitation). We concentrate on short times
and assume a long life time of the holon-doublon pairs. The later decay of the holon-doublon pairs through MM decay \cite{Lenarcic2013,Golez2014,DalConte2015} constitutes another channel
for driving the spin dynamics, which we neglected for simplicity.

\subsection{Two-magnon Raman scattering}
For comparison to the above mentioned absorption mechanisms, we also discuss briefly the two-magnon Raman scattering. Here an incoming photon scatters inelastically in an interaction where two magnons with opposite momenta are created \cite{Fleury1968}. For high $T_c$ cuprate superconductors, the assignment of the spectral weight to two-magnon scattering has been used to determine the values of the leading exchange interactions in the CuO$_4$ planes \cite{Lyons1988}. For CuO, two-magnon scattering was discussed as contributions to the broad peak features around 1100 and 2000 cm$^{-1}$ \cite{Irwin1990}. The fraction of spins which are excited in CuO through two-magnon Raman scattering during a pump pulse can be estimated from the cross-sections known from spontaneous Raman scattering on chemically related copper oxides \cite{Knoll1996}. For YBa$_2$Cu$_3$O$_{6.0}$ it was reported \cite{Knoll1996} a two-magnon Raman cross section $S_{\rm 2M}=1.2\cdot 10^{-3}$ cm$^{-1}{\rm sr}^{-1}$ per unit volume of sample. In the pump probe experiment of \cite{Johnson2012} the pump laser wave length $\lambda=800$ nm corresponds to a photon energy of $E_{\rm phot}=1.55$~eV. The intensity of the pump laser was such that over the 40 fs FWHM pulse lengths, a fluency of $E_{\rm pump}=7$, 14 or 28 mJ/cm$^2$ were incident on the sample. The 4 atom primitive cell of CuO has volume $V_{\rm pc}=0.081$ nm$^{3}$. The total number of photons per unit area impinging the surface of the sample during one pulse is given by $N_{\rm phot}=E_{\rm pump}/E_{\rm phot}$. Multiplying by $S_{\rm 2M}$ and by the volume per Cu, and integrating over $4\pi$ sr we obtain the total number of spin flips created per Cu.
\bgea
\rho&=&4\pi S_{\rm 2M}N_{\rm phot}V_{\rm Cu}=4\pi S_{\rm 2M}\frac{E_{\rm pump}}{E_{\rm phot}}V_{\rm Cu}
\enea
For a fluency $E_{\rm pump}=7$ mJ/cm$^2$ a fraction of two-magnon spin flips $\rho\sim 10^{-8}$ is obtained, a value too small to allow inelastic scattering to be of relevance for driving the spin system from the AF1 to the AF2 phase. Therefore, below we restrict ourselves to the spin dynamics that follows from the absorption of light.

\section{Simulation results}\label{sec:sims}
In this section we present simulations results in which spin system is excited with a varying a fraction of interchanged spin-pairs and photo-induced vacancies, simulating phonon-assisted MM excitation and PD, respectively. In addition, for comparison we simulate the response to sudden changes of the temperature, mimicking phonon-heating. For each case, we have followed the evolution in time of $S(\mathbf{q},t)$. All simulations have been performed with the UppASD software \cite{Skubic2008} for a $36\mathbf{a}\times 4\mathbf{b}\times 36\mathbf{c}$ supercell with periodic boundary conditions. For this cell size $S(\mathbf{q},t)$ has peaks for $\mathbf{q}_{\rm AF1}=(0.5,0,0.5)$ and $S\mathbf{q}_{\rm AF2}=(0.528,0,-0.472)$ and we use the $S(\mathbf{q},t)$ values for these wave vectors to follow the time evolution of the magnetic ordering. We approximate the excitations to be instantaneous, i.e at time $t=0$ we swap a fraction $x$ of pairs of spins or include a fraction $x$ of vacancies in the system. This is a simplification of the actual optical excitation process which takes place for an incident laser pulse that is few tenths of femtoseconds long and the photon-matter interaction developing over multiples of electronic hopping times.

In order to clarify the microscopic origin of the observed dynamics, below we will present simulations with and without coupling to the heat bath. The latter gives undamped ($\alpha=0$) evolution. Under the isotropic Hamiltonian $\mathscr{H}_{\mathrm{iso}}=\mathscr{H}_{\mathrm{exc}}+\mathscr{H}_{\mathrm{bq}}$, both angular momentum and total energy are constants of motion, but energy redistribution between 
different degrees of freedom can still give rise to non-trivial dynamics. At finite damping, $\alpha>0$, energy can dissipate to the environment, enabling the system to relax back to the initial state.

We also note that, despite CuO being an antiferromagnet, at finite temperature there is anyhow a small but finite net magnetic moment, given the circumstance that the thermally fluctuating spins do not fully compensate each other to give identically zero total magnetic moment. 
In equilibrium at $T=160$~K, the magnetic moment in Langevin dynamics $\alpha=0.01$ at times $t<-10$~ps fluctuates with an amplitude $\sim 10^{-3}\mu_{\rm B}$. Using the semi-implicit solver SIB \cite{Mentink2010} for the Landau-Lifshitz equation, the total magnetic moment and its vector components are conserved to $\pm 10^{-7} S$ in $\alpha=0$ evolution over the simulation interval $-10<t<0$~ps. The total energy of the spin system is conserved to $\pm 10^{-4}$ meV over the same simulation time.

The protocol for the simulations has been the following: Starting from the $T=0$~K AF1 configuration the spins are relaxed to be in equilibrium with a finite temperature. The system then evolves for times $-10 < t < 0$~ps, to check that spontaneous AF1-AF2 transitions do not occur, before it is excited at $t=0$ and subsequently evolved to extract the spin dynamics caused by photo-excitation. 

\subsection{Phonon-assisted multimagnon excitation}
In this section results are reported from simulations of phonon assisted MM excitation, which dominates for photon energies below the charge transfer gap. First, we focus on simulations at low temperature $T=10$~K, to show that this causes a nonequilibrium distribution of the spins which are elucidated by histograms of $\{E_i\}$ energies.  Second, we show simulation results for the initial temperature $T=160$~K, from which MM excitation turns out to be sufficient to drive the system in the AF2 phase.

\begin{figure}
\includegraphics[width=\columnwidth]{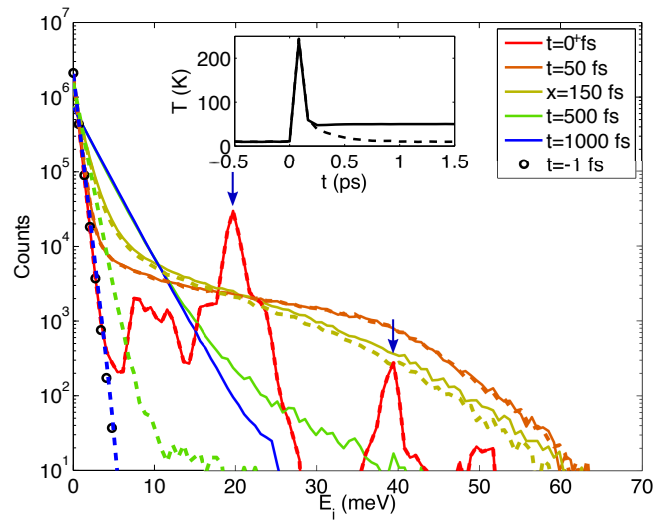}
\caption{\label{fig:energyhistT010} Histogram of energies $\{E_i\}$ after exciting a system in equilibrium at T=10 K (black circles) with a concentration $x=0.02$ of two-magnons flips. The blue arrows indicate the two peaks in the red curve that are corresponding to 2 and 4 magnon excitations respectively. The inset shows the temperature evolution of the spin system. Full (dashed) lines show undamped $\alpha=0$ (damped $\alpha=0.01$) dynamics.}\end{figure}

In Figure~\ref{fig:energyhistT010} is shown the $\alpha=0$ (full lines) and $\alpha=0.01$ (dashed lines) evolution in time of the energy histogram after an excitation of $x=0.02$ (the concentration of \textit{pairs} of spins exchanged) of two-magnon excitations of a spin system initially in thermal equilibrium at $T=10$~K. Upon excitation, energy is injected to the spin system which is brought to a non-equilibrium state. The black symbols display the energy distribution before pump when the spin system is in equilibrium. The red lines for $t=0^+$ fs (after pump) have distinct peaks (marked out with blue arrows) indicating the energy and the intensities of two- and four-magnon excitations. The insets show the evolution of the temperature of the system, with the first few data points displaying least mean square values of the temperature in a non-equilibrium regime in which temperature is not well defined. After $t=1.0$~ps the system has equilibrated to a new temperature of $T=50$~K. We note that this relaxation to a Boltzmann distribution took place in microcanonical evolution. For an isolated spin system this kind of equilibration can occur due to the non-linearities inherit in the Landau-Lifshitz equation, but it is not possible for e.g. for an integrable system like an isolated system of coupled harmonic oscillators.

For the case of evolution $\alpha=0.01$ with finite damping (dashed lines), the excitation energy added to the system dissipates as the system relax to be in equilibrium with the heat bath. For times up to $t=150$ fs, the evolution of the energy distribution is very similar to the system with undamped dynamics. At $t=1.0$~ps, the equilibration is essentially complete and the system has reached a temperature of $T=10.0$~K.

We emphasize that at low temperature distinct peaks corresponding to two- and four-magnon excitation can be seen, indicated by arrows in Figure~\ref{fig:energyhistT010}. Below we study MM excitation at much higher temperatures. In that case, there is already a large fraction of spins that have energies that are of the same magnitude as the energy of a MM excitation event and therefore such sharp features are absent. From this we understand that the average energy given to the spins by MM excitation is in itself temperature dependent. It will take a higher value at low temperatures where the spins are close to perfectly collinear, but a reduced value at intermediate and high temperatures where the effective fields are reduced by thermal fluctuations.

\begin{figure}[t]
\includegraphics[width=\columnwidth]{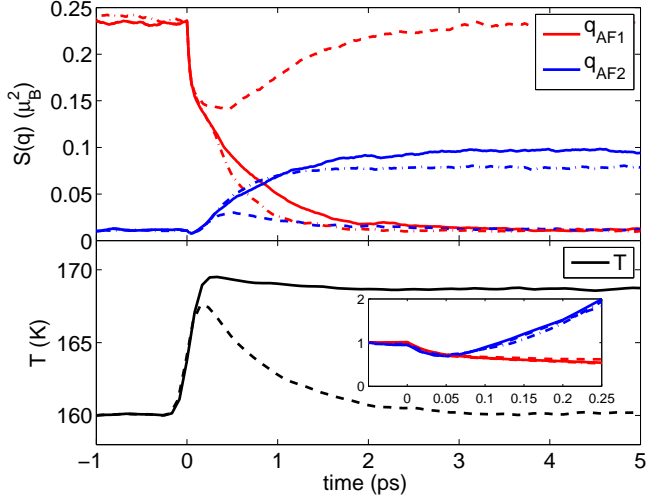}
\caption{\label{fig:ordertemp2MT160}Dynamics following multimagnon excitation. The upper panel shows the trajectory in time of $S(\mathbf{q}_1,t)$ (red) and $S(\mathbf{q}_2,t)$ (blue), averaged over 128 replicas, for $\alpha=0$ (full line) and $\alpha=0.01$ (dashed line) dynamics. $t<0$ is before pump and $t>0$ is after pump. The initial temperature is $T=160$~K, and the initial excitation intensity is $x=0.02$. The lower panel shows the evolution of the temperature of the system. The inset shows the first 250 fs of evolution for the normalised order parameters. The dash-dotted lines are for simulations using the isotropic Hamiltonian $\mathscr{H}_{\mathrm{iso}}$.} 
\end{figure}

The $x=0.02$ excitation did not bring enough energy to push the $T=10$~K system into the AF2 regime, which over few ps relaxed to 50 (10) K for undamped (damped) dynamics. The situation changes if the excitation takes place closer to $T_{\rm N1}$. In Figure \ref{fig:ordertemp2MT160} is shown the results of a simulation where the spin system, initially in thermal equilibrium at $T=160$~K is excited with an excitation intensity of $x=0.02$ and evolved in undamped $\alpha=0$ (full lines) and damped $\alpha=0.01$ (dashed lines) dynamics. For $\alpha=0$ the system undergoes a phase transition from AF1 to AF2 and reaches new equilibrium values of the order parameters after approximately 4 ps. An initial decay can be observed for $S(\mathbf{q}_1,t)$ and also for $S(\mathbf{q}_2,t)$, due to the circumstance that the sublattice magnetization is reduced abruptly in the two-magnon excitation. This can be observed more directly in the inset, in which the normalised quantities $S(\mathbf{q}_1,t)/S(\mathbf{q}_1,t=0)$ and $S(\mathbf{q}_2,t)/S(\mathbf{q}_2,t=0)$ are plotted. We note that within the insets time window of 0.25 ps, the $\alpha=0$ and $\alpha=0.01$ dynamics depart only slightly from each other. As expected the decrease of $S(\mathbf{q}_1,t)$ is monotonous, but also $S(\mathbf{q}_2,t)$ decay over the first 50 fs. For times $t>50$ fs, $S(\mathbf{q}_2,t)$ is increasing as the spin system enters the transition into the AF2 phase. 

\begin{figure}[t]
\includegraphics[width=\columnwidth]{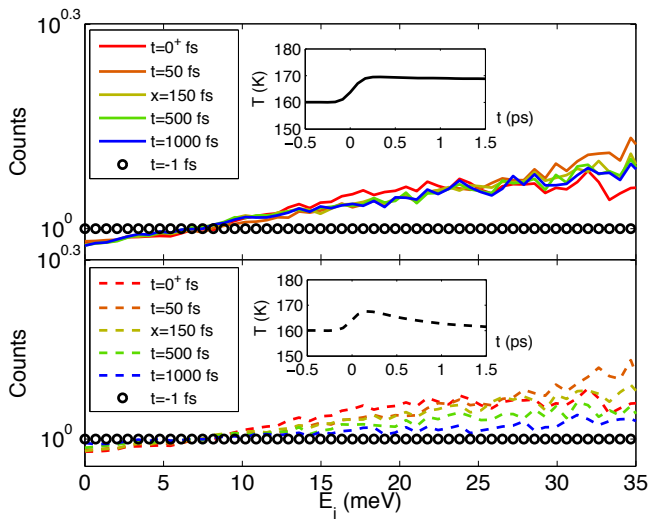}
\caption{\label{fig:energyhistT160}The relative histograms $I(t)/I(t<0)$ of energies $\{E_i\}$ evolving in a) $\alpha=0$ and b) $\alpha=0.01$ dynamics respectively, after exciting a system in equilibrium at T=160 K (black circles) with a concentration $x=0.02$ of two-magnons flips. The quota $I_n(t)/I_n(t<0)$ becomes very noisy at at energies larger than 35 meV, given the overall low bin counts in this energy range. Therefore $I_n(t)/I_n(t<0)$ is shown only for the energy interval 0 to 35 meV. The insets show the temperature evolution of the spin system.} 
\end{figure}

For a system in equilibrium at $T=160$~K, the change to the shape of the energy histogram on two-magnon excitation is much less drastic as compared to the $T=10$~K case shown in Fig.~\ref{fig:energyhistT010}. We visualize the deviation from the thermal equilibrium distribution $I_n(t<0)$ by plotting the relative histogram $I_n(t)/I_n(t<0)$, where $I_n$ is the bin count of the $n$:th bin, as shown in Fig.~\ref{fig:energyhistT160} for $\alpha=0$ [Fig.~\ref{fig:energyhistT160}(a)] and $\alpha=0.01$ [Fig.~\ref{fig:energyhistT160}(b)]. This clearly demonstrates that, similar as in the $T=10$~K case, MM spin flips cause high energy excitations in the spin system. The quota $I_n(t)/I_n(t<0)$ becomes very noisy at at energies larger than 35 meV, given the overall low bin counts in this energy range. Therefore $I_n(t)/I_n(t<0)$ is shown for the energy interval 0 to 35 meV.

\begin{figure}
\includegraphics[width=0.50\textwidth]{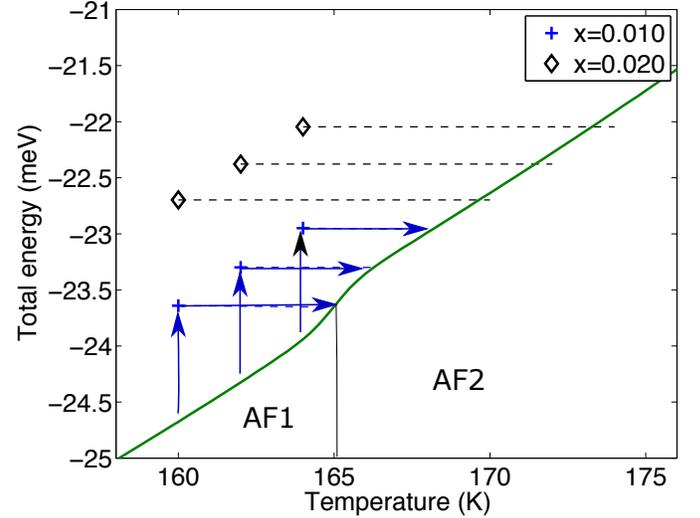}
\caption{\label{fig:TenergyExc4mod}(Color online) The full line shows the energy of the spin system as a function of temperature. The blue and black symbols indicate the initial $T$ and the energy after excitation with strength $x=$0.010 or 0.020 respectively (the concentration of \textit{pairs} flipped) of a system which before pump is in thermal equilibrium at $T=160$, 162 or 164 K.} 
\end{figure}

The lower panel of Figure \ref{fig:ordertemp2MT160} showns the evolution in time of the temperature which for the case of undamped dynamics initially peaks at $T=170$~K. In the subsequent evolution the temperature decreases to $T=169$~K as latent heat is necessary to enter the AF2 phase. As can be seen in Fig.~\ref{fig:TenergyExc4mod}, the energy injected by the initial excitation brings the average spin energy to the value which in thermal equilibrium corresponds to a temperature $T\approx 169$~K. For $\alpha=0.01$ the order parameters $S(\mathbf{q}_1,t)$ and $S(\mathbf{q}_2,t)$ display a transient reduction and enhancement respectively, and the system relaxes back to $T=160$~K within 4 ps. 

In order to analyse how large influence anisotropic terms in the Hamiltonian have on the phase transition, we have pursued simulations for the isotropic Hamiltonian $\mathscr{H}_{\mathrm{iso}}$
. 
The instantaneous excitation of the spin system at $t=0$ in form of swaps of neighboring spins, inject energy but conserve the angular momentum since the swap operator is rotationally invariant. The dash-dotted lines in Fig. \ref{fig:ordertemp2MT160} show the evolution in time of $S(\mathbf{q}_1,t)$ and $S(\mathbf{q}_2,t)$ respectively in simulation of $\mathscr{H}_{\mathrm{iso}}$. Interestingly, we find that the transition between the AF1 phase and the AF2 phase after the initial injection of energy by MM excitation, can proceed solely through redistribution of energy between the isotropic degrees of freedom.

\begin{figure}[t]
\includegraphics[width=\columnwidth]{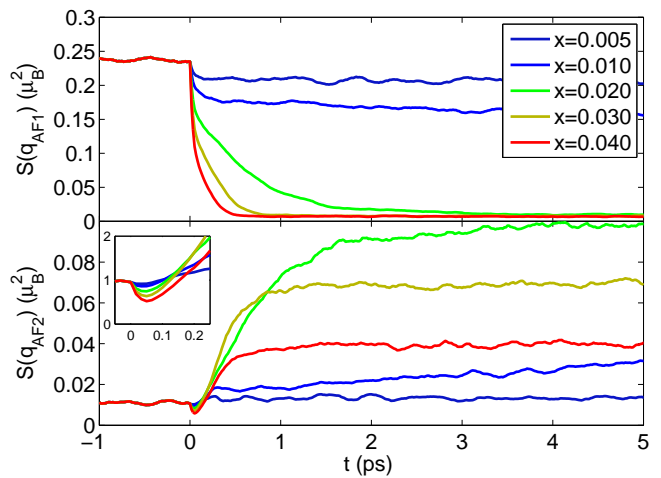}
\caption{\label{fig:order2MT160sweep}Excitation intensity dependence of dynamics following MM excitation. The trajectories in time of $S(\mathbf{q}_1,t)$ (upper panel) and $S(\mathbf{q}_2,t)$ (lower panel), averaged over 128 replicas, and with $\alpha=0$. The initial temperature was $T=160$~K, and the initial excitation intensity was in the range $[0.005,0.040]$. The inset shows the evolution of the order parameters during the first 250 fs.}
\end{figure}

Johnson \textit{et al.} \cite{Johnson2012} investigated the dependence of a switching delay time $t_p^{\rm exp}$, defined as the onset time at which the ratio $S(\mathbf{q}_2,t)/S(\mathbf{q}_1,t)$ would start to grow, on the pump fluency. Starting from lower fluencies $t_p^{\rm exp}$ decreased with increasing fluency, but it was not possible to push it below $t_p^{\rm exp}=400$~fs. In the present model for phonon assisted MM excitation, the analogue of pump fluency is the concentration of initial spin pairs flipped. Results from simulations with initial concentration in the range $[0.005,0.040]$ are presented in Fig.~\ref{fig:order2MT160sweep} and we make the following observations: i) In the interval $100<t<400$~fs, the rate of growth of $S(\mathbf{q}_2,t)$ is very similar in the fluency ranges $[0.010,0.040]$, with $x=0.020$ being the concentration that for $t=5$ ps has given the highest value of $S(\mathbf{q}_2,t)$. For fluencies $x=0.030$ and $0.040$ the spin system is over pumped and brought closer towards the paramagnetic region. ii) The inset displays the evolution for the first 250 fs and reveal that the time $t_p^{\rm sim}$, at which $S(\mathbf{q}_2,t)$ start to grow after the initial dip, takes the value $t_p^{\rm sim}=50$~fs \textit{independent} of fluency. 

For the excitation $x=0.02$ it can be observed in the inset of Fig.~\ref{fig:ordertemp2MT160}, that the onset time $t_p^{\rm sim}$ is not affected by damping $\alpha=0.01$. This can be understood from the following considerations: With the AF1 as reference spin configuration, the highest precession frequencies occur when summing up the exchange interactions to $\sum_i J_{ij}=168$ meV for $f_{\rm max} = \gamma B_{\rm max} = 81.8$ THz, which gives for $\alpha=0.01$ the corresponding frequency of damping motion $f^{\alpha}_{\rm  max}=\alpha f_{\rm  max}=0.818$~THz, or a minimum relaxation time $\tau_{\rm min}=1/(2\pi f^\alpha_{\rm  max})=194$~fs. Thus, damping motion will give significant influence only after hundreds of fs, but will for $\alpha=0.01$ be negligible for the first few hundred fs of evolution, constituting a regime of exchange interaction driven dynamics \cite{Mentink2012,Baryakhtar2013}.

\subsection{Photodoping induced dynamics}\label{sec:pddyn}
The energy of the incident 800 nm laser pump pulse used in \cite{Johnson2012} coincide with the charge transfer (CT) gap of CuO. In this regime one can expect that in addition to the previous mechanism also PD to be an important excitation mechanism.
Photodoping will have two effects. If the incident energy is larger than the CT gap the excited particle hole pairs will relax by the emission of magnon and phonon excitations until they reach the minimum possible unbound CT energy or an excitonic bound state. This transfer of energy can be modelled as a heating of the phononic system or the magnetic system and in our simulations corresponds to a rise of the bath temperature in the damped dynamics. On the other hand, as we will show below, there is a magnetocaloric effect due to the sudden change in the magnetic Hamiltonian. We will analyze this effect in detail neglecting the heating effect, which can be minimized by tuning the incoming laser to the minimum CT energy. The temperature induced dynamics is discussed below.  

As discussed in detail in Ref.~\cite{Hellsvik2014}, in equilibrium one of the main effects of dilution of the spin system is to shift the magnetic phase diagram to lower temperatures. Below we present nonequilibrium studies in which we follow the dynamics after sudden dilution. The concentration of vacancies $x$ is chosen in a similar range as the pairs of interchanged spins discussed with results of phonon assisted MM excitation.
\begin{figure}[b]
\includegraphics[width=\columnwidth]{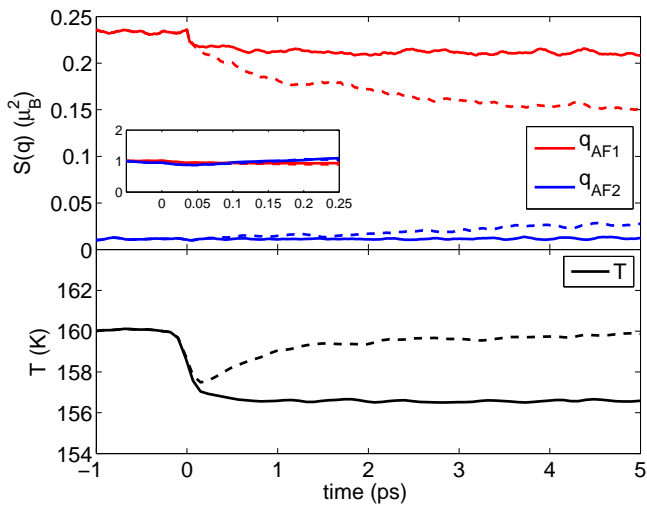}
\caption{\label{fig:ordertempPDT160}Dynamics following on PD. The upper panel shows the trajectory in time of $S(\mathbf{q}_1,t)$, averaged over 128 replicas, for $\alpha=0.00$ (full lines) and $\alpha=0.01$ (dashed lines). The initial temperature was $T=160$~K, and the initial excitation intensity of vacancies was $x=0.02$. The lower panel shows the evolution of the temperature of the system.The inset shows the first 250 fs of evolution for the normalised order parameters, using the same range of the axes as in the inset of Fig.~\ref{fig:ordertemp2MT160}.} 
\end{figure}

\begin{figure}
\includegraphics[width=\columnwidth]{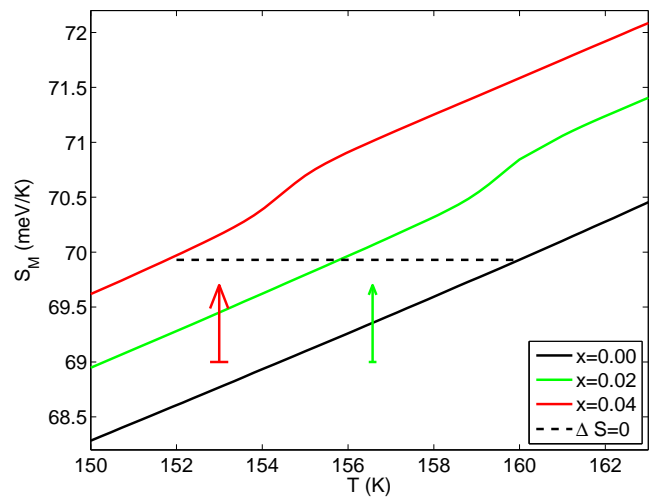}
\caption{\label{fig:Entropy}The entropy $S(T,x)$ for CuO as a function of temperature and vacancy concentration $x$ for pure CuO and for CuO with $x=0.02$ and $x=0.04$. The horizontal dashed line indicates the cooling at constant entropy showing rather close quantitative agreement with the temperatures, indicated by vertical arrows with errorbars, measured within the spin dynamics simulations.}
\end{figure}
\begin{figure}
\includegraphics[width=\columnwidth]{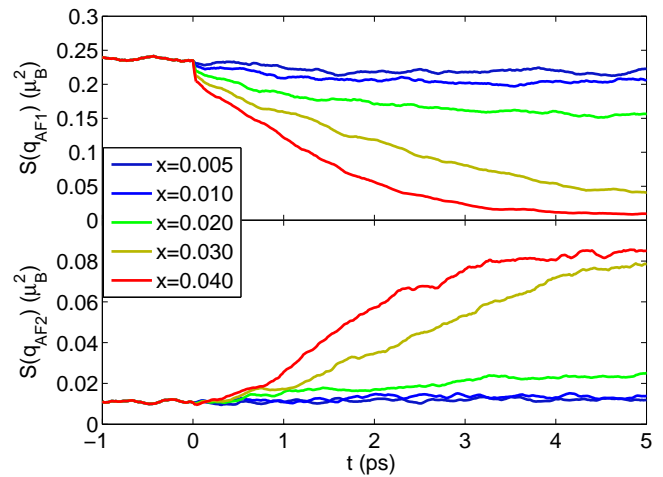}
\caption{\label{fig:orderPDT160sweep}Excitation intensity dependence of dynamics following on PD. The trajectory in time of $S(\mathbf{q}_2,t)$, averaged over 128 replicas, and with $\alpha=0.01$. The initial temperature was $T=160$~K, and the initial excitation intensity of vacancies was in the range $[0.005,0.040]$.}
\end{figure}

In Fig.~\ref{fig:ordertempPDT160} is shown the evolution of the order parameters $S(\mathbf{q}_1,t)$ and $S(\mathbf{q}_2,t)$ following on PD that introduces $x=0.02$ vacancies. Full lines show evolution at $\alpha=0$. The $S(\mathbf{q}_1,t)$ ordering experience an abrupt step as the sublattice moment is reduced, correlated with the reduced number of spins in the system. After this initial drop, $S(\mathbf{q}_1,t)$ levels out to a slightly reduced value and $S(\mathbf{q}_2,t)$ is, over the time scale of 5 ps, not affected. Interestingly, the temperature of the system experiences a rapid drop from $T=160$~K to 157 K. We explain this drop as an ultrafast magnetocaloric effect, where the role of the external magnetic field is replaced by the change of the effective exchange field on the spins by photo-doping.

To substantiate our interpretation of a magnetocaloric effect we computed the equilibrium entropy per spin as a function of $T$ for different vacancy concentrations $x$,  $S(T,x)=S_0(x)+ \int_{T_0}^T\!d\tau\,{C_\text{M}(\tau\!,x)}/{\tau}$ and the magnetic heat capacity $C_M$ is computed from the energy fluctuations $\langle (\delta E)^2\rangle$. Here the total entropy is divided by the actual number of spins in the sample, not the number of sites. To obtain an accurate account of the entropy differences, we matched $S_0(x)$ to $S_0(0)$ in the high-temperature paramagnetic phase at $T=1260$~K, while $T_0=1$~K.

The curves shown in Fig.~\ref{fig:Entropy} are qualitatively similar to what would be obtained for independent spins in a uniform magnetic field, with the upper curves corresponding to lower magnetic field. The reason is that both the magnetic field and the interaction constrain the spin motion, effectively reducing the fluctuating phase space and lowering the entropy. Taking a reference spin, after a sudden quench of the magnetic field (classical magnetocaloric effect) or after a neighbor is eliminated (present simulation) the spin finds itself in a state which is much more ordered than what it should be at the initial temperature. This corresponds to the adiabatic jump shown in Fig.~\ref{fig:Entropy} with the dashed line. The vertical arrows show the temperatures obtained in spin dynamics simulations after internal equilibration in $\alpha=0$ dynamics, with the error bars stemming from the temporal fluctuations and the measurement of the temperature (see Appendix \ref{sec:tempmeas}). We can observe that in the spin dynamics simulations the temperatures are lowered almost but not quite as much as the constant entropy jump implies.

Similar rapid drops in the temperature are obtained with changing the strength of the Heisenberg exchange interaction $J_z\rightarrow (1-x)J_z$ (data not shown). In this case we can estimate the effective field seen by a single spin simply from $zxJ_zS^2=g\mu_\text{B}SB_\text{eff}$, with $J_z$ the dominant Heisenberg exchange interaction \cite{Hellsvik2014}, yielding $B_\text{eff}=10$~T at $x=0.01$ and an efficiency of 0.15~K/T. This is comparable but slightly lower than the efficiencies reported in \cite{Ma2014} and references therein, which is understandable since our results are obtained deep in the ordered phase of CuO.

Although the system thermalizes even at $\alpha=0$, $S(\mathbf{q}_2,t)$ is not affected on the time scale of the simulation. Hence, despite the fact that the introduction of vacancies reduces $T_{N1}$, our simulations do not show the evolution to a different phase. We can understand this since the shift in $T_{N1}$ is smaller than the magnetocaloric effect. Indeed, already at $x=0.02$ we obtain a reduction to $T\approx 156$~K from the magnetocaloric effect, while the lowering of $T_{N1}$ to a similar value requires a higher vacancy concentration, $x\approx0.03$ \cite{Hellsvik2014}. This observation implies that the phase transition due to the lowering of $T_{N1}$ is possible when the system is kept in contact with the bath at the original temperature $T=160$~K after the introduction of vacancies. To confirm this, we have carried out simulations at $\alpha=0.01$, shown by dashed lines in Fig.~\ref{fig:ordertempPDT160}. In this case $S(\mathbf{q}_1,t)$ is further reduced after the fast initial drop, while $S(\mathbf{q}_2,t)$ increases. For vacancy concentration $x=0.02$, the growth of $S(\mathbf{q}_2,t)$ is still modest over the first ps, probably because the system is still to close to $T_{N1}$ and the transition is broadened due to finite size effects. As shown in Fig.~\ref{fig:orderPDT160sweep}, a stronger response can be seen for doping up to $x=0.04$, for which also $S(\mathbf{q}_1,t)$ decays completely and we are thus able to see the full transition on the time scale of our simulation. These results demonstrate that, opposed to the case of phonon-assisted MM excitation, the timescale at which the transition can occur is limited by the coupling to the bath and is therefore slower. 

\subsection{Temperature induced dynamics}\label{sec:tdyn}
\begin{figure}[b]
\includegraphics[width=\columnwidth]{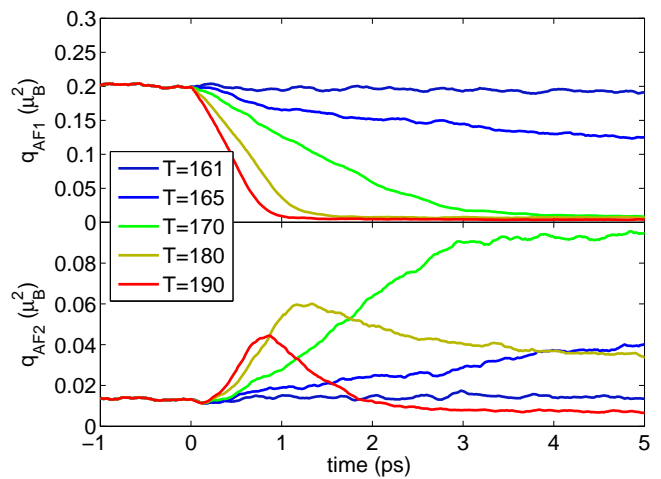}
\caption{\label{fig:orderHeatSweep}Dynamics following on a change of the heat bath temperature. The trajectory in time of $S(\mathbf{q}_1,t)$ and $S(\mathbf{q}_2,t)$, averaged over 128 replicas, and with $\alpha=0.01$. The initial temperature was $T=160$~K.}
\end{figure}

For comparison, we also simulated direct heat-induced spin dynamics, mimicking a rapid rise of the phonon temperature by laser absorption. One can also take the bath to be excited e-h pairs in which case, in first approximation, this effect should be added to the one of the PD computed above. The response of the spin system to an abrupt change of temperature at $t=0$ is shown in Fig.~\ref{fig:orderHeatSweep}. For simplicity, we consider a stepwise change of temperature which should give an indication of how fast the system react to heating. Unlike the case of phonon assisted MM excitation and PD, there is no instantaneous drop in the sublattice magnetization, so that $S(\mathbf{q}_1,t)$ and $S(\mathbf{q}_2,t)$ are continuous at $t=0$. This absence of instantaneous decrease in the structure factor is in contrast with the diffraction data in \cite{Johnson2012}. We conclude that this mechanism is not playing an important role in those experiments. 

The speed of the response to heating is parametrically dependent on the strength of the damping parameter $\alpha$ \cite{Hellsvik2008}. Among the set of temperatures included in Fig.~\ref{fig:orderHeatSweep}, the highest value of $S(\mathbf{q}_2,t)$ occur for $T=170$~K. Unlike the case of phonon assisted MM excitation, the rate with which $S(\mathbf{q}_2,t)$ grows, is increasing for stronger pump fluence.

\section{Conclusions}\label{sec:conc}
In summary, we have explored how excitation with femtosecond pulses of near infrared laser light can drive a phase transition in CuO. On the basis of atomistic spin dynamics simulations we have found that photo-doping causes an ultrafast cooling of the spin system which can be understood as a magnetocaloric effect, where the collinear antiferromagnetic order drops rapidly and the spiral phase grows at a much slower rate due to the photo-induced reduction of $T_{N1}$. Phonon-assisted MM excitation increase the temperature of the spin system on sub-picosecond timescales and a rapid drop of both collinear and spin spiral order is found, while only the spiral order grows gradually after this rapid drop. The phonon-assisted MM excitation is found to cause the fastest phase transition and we argue that this is related to the fact that only for this case the phase transition can proceed by pure exchange-driven spin dynamics \cite{Mentink2012,Baryakhtar2013} alone. In addition, estimates of the quantum mechanical efficiency of inelastic light scattering was found to be too low to play any significant role in driving the phase transition, while a purely temperature driven dynamics cannot reproduce the initial abrupt reduction of the order parameters observed in \cite{Johnson2012}.

Finally, by comparing our results with experiments, we observe that only phonon-assisted MM absorption captures the experimentally observed saturation with the pump fluence and the simultaneous initial drop of both collinear and spin-spiral order. On the other hand, in the experiments the timespan over which this simultaneous drop takes place is a factor of 8 larger than in our simulations. There are several reasons for this discrepancy. First, the simulations assume an instantaneous excitation, whereas experimental laser pulses in \cite{Johnson2012} had a FWHM maximum of 40 fs. Moreover, in the experiment the laser pump pulse coincided with the band gap of the material, wherefore both PD decay and phonon assisted MM excitation are possible excitation mechanisms. On the level of ASD simulations, it is not straightforward to realistically model of the interplay between these processes. In addition, we assume that concentration of vacancies remains constant on the timescale considered, while  the eventual decay of the photo-excited holon-doublon pairs can also proceed through MM excitation \cite{Lenarcic2013,Golez2014,DalConte2015}.
Such decay can therefore drive the spin dynamics across the phase transition in a similar fashion as phonon-assisted MM excitation. Our simulations also show that these processes are not captured by a simple increase of the bath temperature. Similar to the absorption of photons we expect that an initial state with MM excitation captures this effect more realistically. 
Clearly, modeling the simultaneous charge and spin dynamics goes beyond the semi-classical approach employed here and would be very interesting to pursuit in future investigations.  
It would be also interesting to repeat the experiments varying the laser wave-length to the middle-IR region to verify our predictions in the pure MM regime. 

\acknowledgments
We acknowledge fruitful discussions with D. Bossini, M. Eckstein, A. V. Kimel, R. V. Mikhaylovskiy, and S. Bonetti. J. H. is supported by the Swedish Research Council. J. H. M acknowledges funding from the Nederlandse Organisatie voor Wetenschappelijk onderzoek (NWO) by a Rubicon and VENI Grant and from the European Research Council (ERC) Advanced Grant No. 338957 FEMTO/NANO. J.L was partially supported by the Italian MIUR under project PRINRIDEIRON-2012X3YFZ2. We acknowledge computational resources and support from CINECA and from SNIC.

\appendix

\section{Measuring temperature}\label{sec:tempmeas}
In this appendix we compare two different methods to sample the temperature of a spin system: the method developed by Ma \textit{et al.}~\cite{Ma2010} and our own construction based on energy histograms. The former has the advantage that a closed expression is used to measure the spin temperature (Eq. 16 of Ref.~\cite{Ma2010}, in the following referred to as the MDSW equation), whereas an advantage of the latter is that the histograms also indicate how long it takes before the spin system is in internal equilibrium as signified by deviations from a Boltzmann distribution.

\begin{figure}
\includegraphics[width=0.50\textwidth]{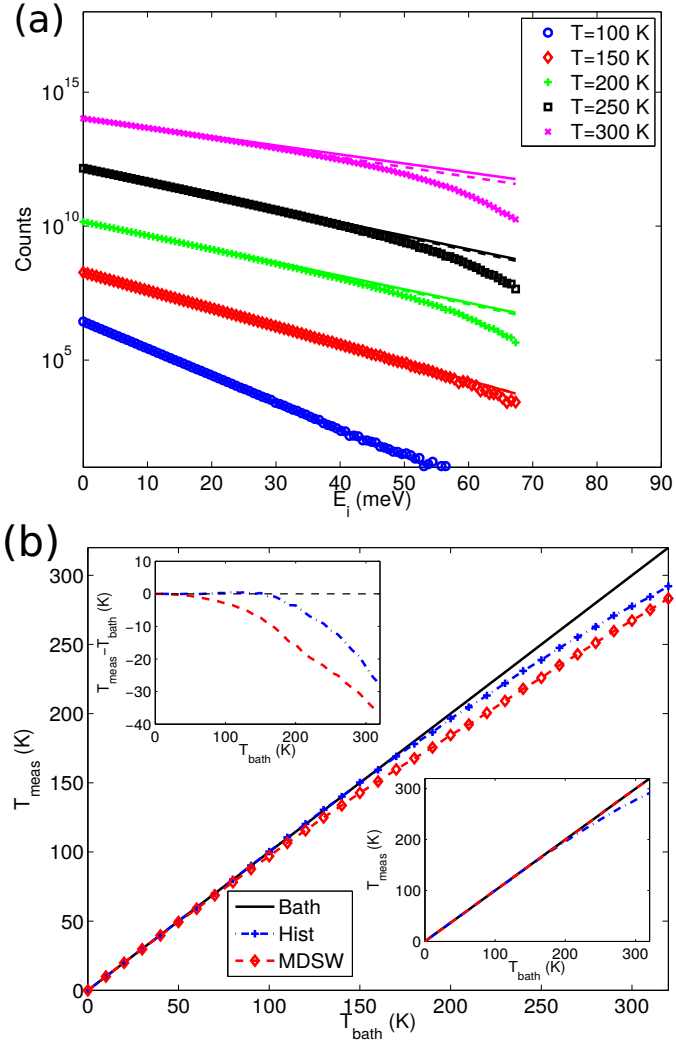}
\caption{\label{fig:energyfits6B} (a) Histograms (symbols) of energies $E_i$ for spin systems in thermal equilibrium. The data for different temperatures have been shifted along the $y$-axis. The full lines show the Boltzmann distribution for the temperature of the heat bath, dashed lines show the fitted temperatures. (b) The measured spin system temperature obtained with the histogram method (blue curve) or the MDSW formula (red curve). The upper inset displays the deviation of the measure temperatures to the the bath temperature. The lower inset shows the measured spin system temperature for the bilinear in spin Hamiltonian $\mathscr{H}_{\mathrm{bilin}}$.} 
\end{figure}

In Figure~\ref{fig:energyfits6B}(a) are shown in log-linear graphs the histograms for $T=100, 150, 200, 250$, and 300 K. For $T=150$~K the bin count deviate from a straight line only for the highest energies, and the histogram can be fitted to a Boltzmann distribution with temperature $T_{\rm fit}=150\pm0.5$~K. At $T=200$~K the deviation is more pronounced and the fitted temperature is slightly underestimated, $T_{\rm fit}\approx 196$~K. At higher temperatures the deviation is even more pronounced.

For the bilinear in spin Hamiltonian $\mathscr{H}_{\mathrm{bilin}}=\mathscr{H}_{\mathrm{exc}}+\mathscr{H}_{\mathrm{ani}}$, i.e. omitting the biquadratic interaction $\mathscr{H}_{\mathrm{bq}}$, the MDSW equation measures the temperature accurately also in the paramagnetic phase, as shown in the lower inset of Fig.~\ref{fig:energyfits6B}(b). However, for our full Hamiltonian $\mathscr{H}_{\mathrm{M}}$ the histogram method works better and does at a $T_{\rm bath}=170$~K undershoot with only 1 K. The deviations of the two methods from the heat bath temperatures are displayed in the upper inset of Fig.~\ref{fig:energyfits6B}(b). For the simulations discussed in Sec.~\ref{sec:sims}, temperatures are in the range $10<T_{\rm fit}<170$~K and the histogram method is expected to be sufficiently accurate. 

\end{document}